\documentclass[aip,jcp,amsmath,amssymb,
reprint]{revtex4-1}
\usepackage{graphicx}
\usepackage{dcolumn}
\usepackage{bm}
\usepackage{array,mathtools,amssymb,booktabs}
\usepackage{relsize}
\usepackage{lmodern} 
\usepackage{upgreek}
\usepackage[hidelinks]{hyperref}

\begin{document}
\title{Sequence Dependence of Critical Properties for 2-letter Chains}

\author{Athanassios Z. Panagiotopoulos}
\email[email:]{azp@princeton.edu}
\affiliation{Department of Chemical and Biological Engineering, Princeton University, Princeton, NJ 08544, U.S.A.}
\date{\today} 

\begin{abstract}
Histogram-reweighting grand canonical Monte Carlo simulations are used to obtain the critical properties of lattice chains composed of solvophilic and solvophobic monomers.  The model is a modification of one proposed  by Larson \emph{et al.} [J. Chem. Phys. 83, 2411 (1985)], lowering the ``contrast'' between beads of different type to prevent aggregation into finite-size micelles that would mask true phase separation between bulk high- and low-density phases. Oligomeric chains of length between 5 and 24 beads are studied. Mixed-field finite-size scaling methods are used to obtain the critical properties with typical relative accuracies of better than $10^{-4}$ for the critical temperature and $10^{-3}$ for the critical volume fraction. Diblock chains are found to have lower critical temperatures and volume fractions relative to the corresponding homopolymers. Addition of solvophilic blocks of increasing length to a fixed-length solvophobic segment results in a decrease of both critical temperature and critical volume fraction, with an eventual slow asymptotic approach to the long-chain limiting behavior. Moving a single solvophobic or solvophilic bead along a chain leads to a minimum or  maximum in the critical temperature, with no change in critical volume fraction. Chains of identical length and composition have a significant spread in their critical properties, depending on their precise sequence. The present study has implications on understanding biomolecular phase separation and for developing design rules for synthetic polymers with specific phase separation properties. It also provides data potentially useful for the further development of theoretical models for polymer and surfactant phase behavior. 

\end{abstract}

\maketitle

\section{Introduction}

The Monte Carlo method for computational determination of condensed-matter properties was introduced in the seminal paper by Metropolis \emph{et al.} \cite{Met53} 70 years ago, for determination of the equation of state of hard disks. In the intervening period, the method has formed the basis for many major developments in statistical mechanics of fluids and solids. The original method was designed for sampling configurations in the canonical ($NVT$) ensemble. Development of Monte Carlo  algorithms for the constant-pressure\cite{Woo68} ($NPT$) and grand canonical\cite{Nor69} ($\mu VT$) ensembles took place in the late 1960's and greatly facilitated the study of phase transitions. Free-energy calculation methods for crystal phases became available in the early 1980's\cite{Fre84} and made possible the determination of solid relative stabilities and their solubilities in liquids. The Gibbs ensemble Monte Carlo method for direct determination of fluid-phase equilibria\cite{Pan87} and efficient histogram-reweighting algorithms\cite{Fer89} were developed in the late 1980's. 

Systems of interest for the present study are relatively short chains composed of two different bead types, solvophilic and solvophobic. It is well-established that monomer sequence has a significant impact on chain phase and aggregation behavior. While early work on such systems focused on modeling surfactant solutions,\cite{Lar85,Lar88,Flo99} recent interest in the effect of chain sequence on the phase behavior has been sparked by the discovery of the important role that liquid-liquid phase separation plays in cell biology.\cite{ Bra09,Ban17}  Intrinsically disordered proteins are often key determinants of biomolecular phase separation, with the observed behavior being highly sensitive to their overall composition and precise amino acid sequence.\cite{Das20,Ran24} There is also significant interest in the properties of synthetic polymers with well-defined blocks of solvophilic and solvophobic segments.\cite{Des24,Tay24} The use of simple, coarse-grained models with only a handful of independent control parameters helps provide a fundamental understanding of the underlying physical phenomena by  excluding complications arising from the presence of many different monomer and interaction types present in real chemical or biological systems.

Relevant prior work in the area of sequence-dependence of critical properties for coarse-grained chain models includes the study of Statt \emph{et al.}\cite{Sta20} for a continuum-space 2-letter heteropolymer, using direct coexistence (interfacial) molecular dynamics simulations. A number of different 20-mer chains were studied, yielding a range of possible dense-phase morphologies and a clear dependence of the critical temperature on a ``blockiness'' parameter. A limitation of that work was the use of interfacial simulations for determination of critical parameters, which results in only approximate values. This limitation was overcome in a subsequent paper\cite{Ran21} that utilized a  2-letter lattice model and grand canonical Monte Carlo simulations. That study provided accurate critical parameters for the model chains, as well as a path for distinguishing clearly between true phase separation and aggregation into finite-size clusters. The scaling of chain critical temperature and volume fraction with respect to a normalized metric for chain blockiness was also determined.  In another recent study,\cite{Pan23} the same 2-letter lattice chain model was used to shed light on the interplay between phase separation into macroscopic (bulk) fluid phase on one hand, and formation of finite-size micelles on the other. It was found that certain triblock sequences with solvophobic ends can undergo both micellization and macroscopic phase separation.  

An element missing from these prior studies that I plan to address in the present work is the quantitative dependence of the critical parameters on overall chain length, as well as their dependence on the size of the solvophilic and solvophobic chain segments. In addition, as explained in the ``Model'' subsection, a modification of the 2-letter chain model parameters was implemented to prevent aggregation into finite-size micelles for all sequences studied and to restore symmetry with respect to the phase behavior for chains consisting of a single bead type. The overall aim of the study is to provide accurate data that could be used for the development of theoretical or machine-learning methods for describing the effects of sequence and chain length on phase separation for synthetic or biological polymers. 

\section{Model and Methods}

\subsection{Model}

The model used in this work is a modification of the  lattice surfactant model first proposed by Larson \emph{et al.} \cite{Lar85,Lar88} The model consists of linear chains of $r$ beads connected by bonds, each bead occupying a single site on a simple cubic lattice. The range of chain lengths studied here was from $r=5$ to $r=24$. The use of shorter chain lengths restricted the size of sequence space and ensured that there was sufficient sensitivity of the critical parameters to sequence variations involving even a single monomer moving along the chain.

Bonds between adjacent beads in the model chains can be in the [0 0 1], [0 1 1], and [1 1 1] directions of the lattice, as well as their  allowable rotations and reflections, resulting in 26 total possible connectivity vectors. There are two bead types, solvophobic and solvophilic, conventionally labeled as ``T'' and ``H'', respectively, for ``tail'' and ``head'' groups in the original surfactant-oriented incarnation of the model. Non-bonded beads interact along the 26 possible lattice connectivity directions. There are no interactions with empty lattice sites, assumed to contain a monomeric solvent.  Thus, the model has three independent energy parameters, specifically $\epsilon_{\mathrm{TT}}$, $\epsilon_{\mathrm{HT}}$, and $\epsilon_{\mathrm{HH}}$. In the original version of the model,\cite{Lar85,Lar88}  the values of the parameters were taken to be $\epsilon_\mathrm{TT}=-2$ and $\epsilon_\mathrm{HT}=\epsilon_\mathrm{HH}=0$. The unit of energy also sets the temperature scale for the model.  Subsequent studies \cite{Ran21,Pan23} used the trivial change $\epsilon_\mathrm{TT}=-1$, $\epsilon_\mathrm{HT}=\epsilon_\mathrm{HH}=0$, which simply rescales the temperature by a factor of 2. With this choice of model parameters, implying a strong contrast between solvophobic and solvophilic interactions, many sequences were found to form micellar finite-size aggregates instead of macroscopic bulk liquid phases\cite{Ran21} at the range of temperatures over which sampling of equilibrium states is feasible. This is an undesirable feature for the present study, because it prevents a direct comparison of critical parameters for many candidate sequences. In order to overcome this limitation, a modified parameter set with reduced contrast between solvophobic and solvophilic beads is utilized here, as follows:
\begin{equation} \label{eq:param} 
		\epsilon_\mathrm{TT} =  -\frac 3 4 \ \ \  ;  \ \ \  \epsilon_\mathrm{HT} = \epsilon_\mathrm{HH}  = -\frac 1 4 
\end{equation}

The relative strength of attractions between solvophobic and solvophilic beads is now 3 (rather than $\infty$ in the previous version of the model), a factor that will be used to normalize critical temperatures with respect to composition variations. This modification also restores symmetry of behavior for chains at the limit at which they consist of a single monomer type. Thus, solvophilic H$_r$ chains now have the same behavior as solvophobic T$_r$ chains, modulo a factor of 3 rescaling energies, chemical potentials, and temperatures. In the previous version of the model, purely solvophilic chains are repulsive and do not undergo a phase transition into a condensed liquid phase at any finite temperature. 

\subsection{Computational Methods}

The computational methodologies used in the present work are similar to those of Ref. \citenum{Pan23} and will only be briefly summarized here for completeness. Monte Carlo simulations in the grand canonical ($\mu VT$) ensemble in cubic boxes of edge length $L$ were used to obtain all aspects of thermodynamic behavior for the  model. The volume fraction when $N$ chains of length $r$ are present in the simulation box is defined as $\phi=Nr/L^3$. To facilitate insertion and removal of chains, an athermal Rosenbluth algorithm\cite{Ros55} was implemented, as detailed in Ref. \citenum{Pan98}. Source codes, example input and output files, and a list of the runs performed for each system are available online, as explained in the  Data Availability Statement at the end of this article. Statistical uncertainties were obtained from the standard deviation of results from four independent runs at identical conditions, each with a different random number seed. 

Histogram reweighting with the Ferrenberg-Swendsen algorithm\cite{Fer89} was used to rescale information from runs at one set of thermodynamic conditions to another and to combine runs that have a reasonable degree of overlap in the number of particles $N$ and energy $E$ covered in each. The combined histograms were then used to obtain distributions of particle numbers and energy, $P(N,E)$, as a continuous function of $\mu$ and $T$. To determine critical points, the mixed-field finite-size scaling method of Bruce and Wilding \cite{Bru92,Wil92} was used. The critical temperature $T_\mathrm{c}$, critical chemical potential, $\mu_\mathrm{c}$, and number-energy field mixing parameter, $s$, were optimized simultaneously, using histogram data near the expected critical point, by minimizing deviations between the observed order parameter distribution and the 3-dimensional Ising universal curve obtained as an analytical expression by Tsypin and Bl\"ote.\cite{Tsy00}  Phase coexistence curves (binodals) were determined from the condition of area equality for the low-$N$ and high-$N$ regions of the $P(N)$ distributions. 

For most chain architectures studied, the simulation box length was set at $L=15$ in lattice units. For systems of chains with $r>8$, larger boxes of $L=20$ and $L=25$ were used to ensure that the box dimensions were kept much bigger than chain average dimensions. These simulation box lengths resulted in average number occupancies of between 100 and 300 chains at the critical temperature and chemical potential. The choice of system size is a compromise between using smaller systems (which enable significantly faster calculations) and matching the universal order parameter distribution closely (which only happens at sufficiently large systems). There are systematic effects of system size on the critical parameters,\cite{Ork99} but for the systems studied here these are generally much smaller than the differences due to varying chain architectures that are of primary interest for the present work. In addition,  comparisons here are made for similar system sizes across architectures and thus would not substantially change if the critical parameters were to be systematically extrapolated to infinite simulation system size, at significant additional computational cost. 

\section{Results and Discussion}

\subsection{Chain-length Dependence}

\begin{figure}
    \centering
    \includegraphics[width=3.5 in]{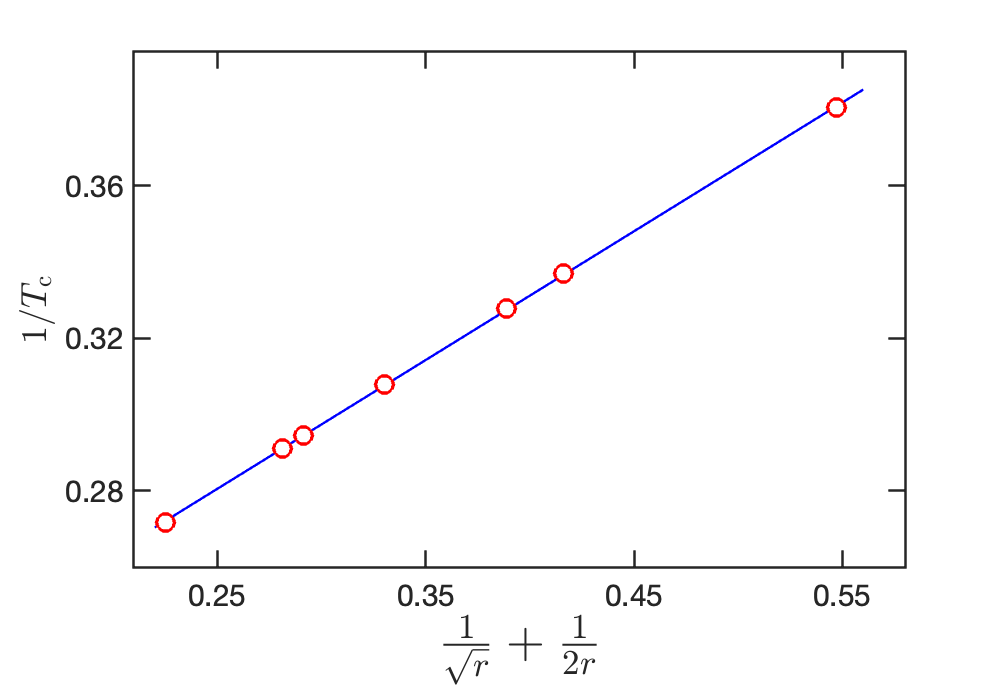}
    \caption{Inverse critical temperature, $1 / T_\mathrm{c}$ as a function of the parameter $ {\frac 1 {\sqrt r}  + \frac 1 {2r} }$ , for homopolymers of type H$_r$. For homopolymers of type T$_r$, temperature will be higher by a factor of exactly 3. Points (from right to left) correspond to $r=5,8,9,12,15,16$, and 24.  The line is a linear-least-squares fit to the data. Statistical uncertainties are smaller than symbol size.} 
 \label{fig:Tc0}
 \end{figure}

The first task of interest was to establish the dependence of the critical temperature and volume fraction on chain length for homopolymer sequences between $r=5$ and $r=24$, matching the range of chain lengths of interest. Results for the critical temperature are shown in Fig. \ref{fig:Tc0}. The figure is for ``all solvophilic'' H$_r$ sequences. For ``all solvophobic'' T$_r$  sequences, the critical temperature would be scaled up by a factor of 3, following the ratio of interaction parameters $ \epsilon_\mathrm{TT} /  \epsilon_\mathrm{HH} $ from Eq. \ref{eq:param}. These critical temperatures are in excellent agreement with prior results for $r=8$ and $r=16$ using the same model.\cite{Pan98} According to Flory-Huggins theory\cite{Flo53}, the inverse critical temperature, $1 / T_\mathrm{c}$, is expected to scale as:
\begin{equation} \label{eq:Tc} 
		\frac 1 {T_\mathrm{c}(r)} - \frac 1 {T_\mathrm{c}(\infty)} \propto  {\frac 1 {\sqrt r}  + \frac 1 {2r} } 
\end{equation}
As can be seen in the figure, this scaling is followed quite closely, as also seen for a broader range of chain lengths in Ref. \citenum{Pan98}.  The extrapolated value is $T_\mathrm{c}(\infty)=20.4\pm0.1$, a bit lower than the estimate in\cite{Pan98} which not surprising, given the lower range of $r$ values examined here. 

\begin{figure}     
    \centering
    \includegraphics[width=3.5 in]{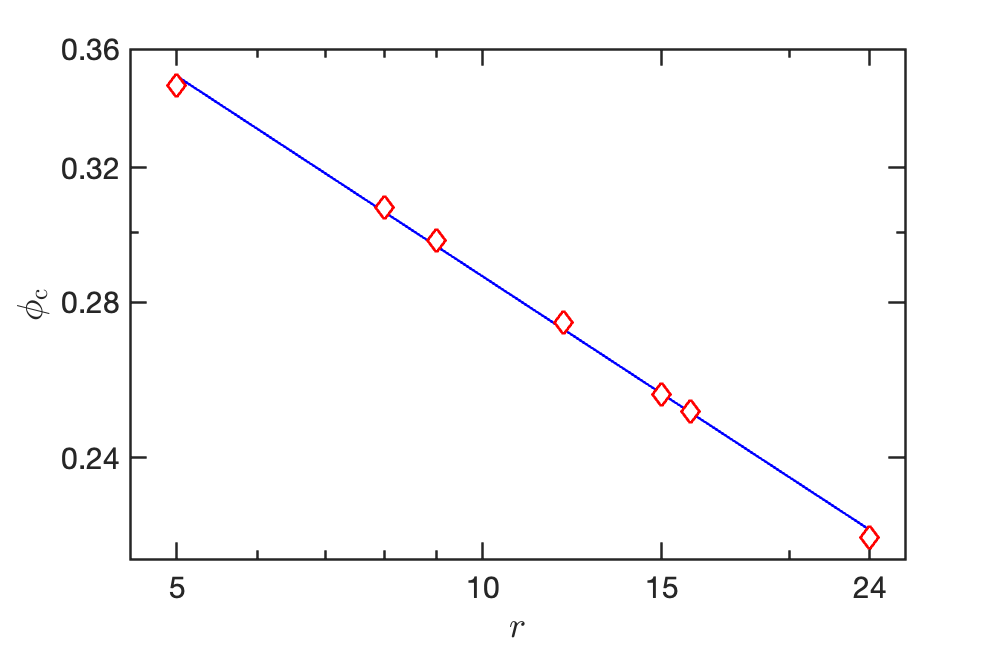}
    \caption{Critical volume fraction $\phi_\mathrm{c}$ as a function of the chain length $ r $, on a log-log scale, for homopolymers H$_r$ or T$_r$ of length between $r=$5 and $r=$24. The line is a linear-least-squares fit to the data. Statistical uncertainties are comparable to symbol size.} 
\label{fig:phic0}
\end{figure}

The dependence of the critical volume fraction on chain length for homopolymers is shown in Fig. \ref{fig:phic0}. The critical volume fraction for homopolymers does not depend on the energy scale, so it is identical for H$_r$ and T$_r$ chains. In agreement with Flory theory\cite{Flo53}, it is seen that the logarithm of the critical volume fraction depends linearly on the logarithm of chain length. The slope obtained from the data is $-0.29\pm0.01$, which is again lower than the exponent obtained in\cite{Pan98} by fitting the range of chain lengths of $64\le r \le1000$. The chain lengths studied here are too short to match the asymptotic long-chain length behavior for the scaling of the critical parameters, but the reason of obtaining these relationships is to obtain analytical expressions for the critical temperature and critical volume  over the chain lengths of interest, which in turn are used in  comparisons when chain composition is varied as well as the chain length.

\subsection{Block-length Dependence for Diblock Chains}

\begin{figure}   
    \centering
    \includegraphics[width=3.3 in]{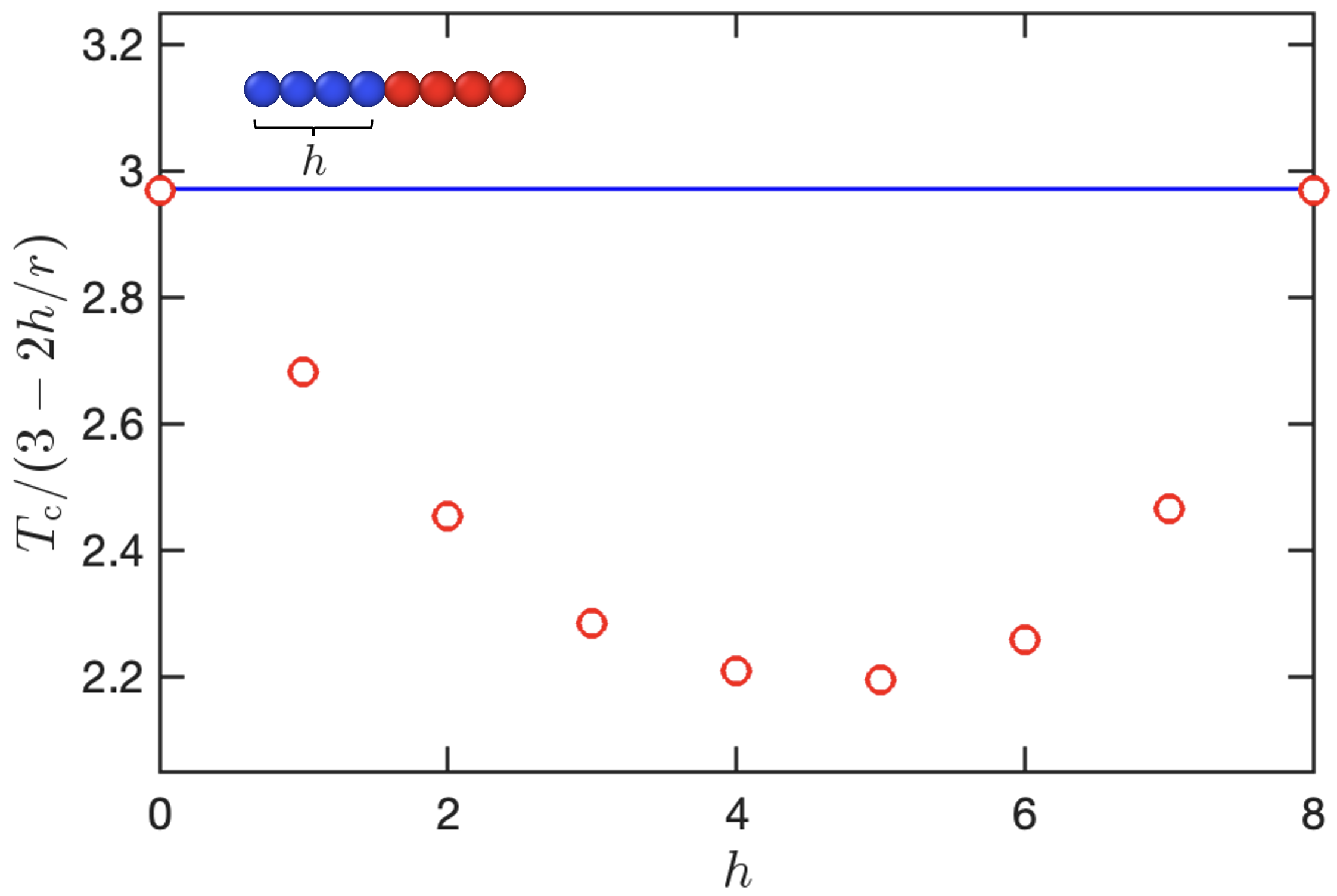}
    \caption{Normalized critical temperature, $T_\mathrm{c} / (3-2h/r)$, as a function of the solvophilic segment length $h$, for diblock chains H$_h$T$_{r-h}$ with overall length $r=8$. The horizontal line is the expected value based on the chain length. Statistical uncertainties are smaller than symbol size.} 
 \label{fig:Tc_db}
 \end{figure}

The next task was to examine the dependence of model critical parameters on the length of the solvophilic segment, $h$, for chains of fixed overall length of 8 beads, as shown schematically in the inset to Fig. \ref{fig:Tc_db}. The overall architecture is H$_h$T$_{r-h}$, with $r=8$. For the full-contrast parameter set studied in Ref \citenum{Pan02}, chains with $h$ between 4 and 7 were shown to micellize, but the reduced-contrast interaction set used here ensures that all chain lengths undergo normal phase separation. This is confirmed by the match of the order parameter distribution to the 3-dimensional Ising universality class, following the procedure of Ref. \citenum{Pan23}.  One complication that arises in comparing the behavior of different architectures is that chains of variable $h$ have different compositions. This can be taken into account by normalizing the temperature as $T_\mathrm{c} / (3-2h/r)$, where the factor of 3 is the ratio of  $ \epsilon_\mathrm{TT} /  \epsilon_\mathrm{HH} $ from Eq. \ref{eq:param}. The expression linearly interpolates between the  limits for the critical temperatures of the homopolymers, T$_8$ ($h=0$) and H$_8$ ($h=r$).  As seen in Fig.  \ref{fig:Tc_db}, the normalized critical temperature shows a minimum at $h=5$, with the corresponding value 26\% lower than for the homopolymer, a rather large effect. This is physically the result of the solvophilic H beads ``screening'' interactions between solvophobic T beads, thus making them effectively weaker. The asymmetry of the behavior (skewed towards higher $h$ values) results from the greater efficiency of this screening by longer H segments. 

\begin{figure}
    \centering
    \includegraphics[width=3.5 in]{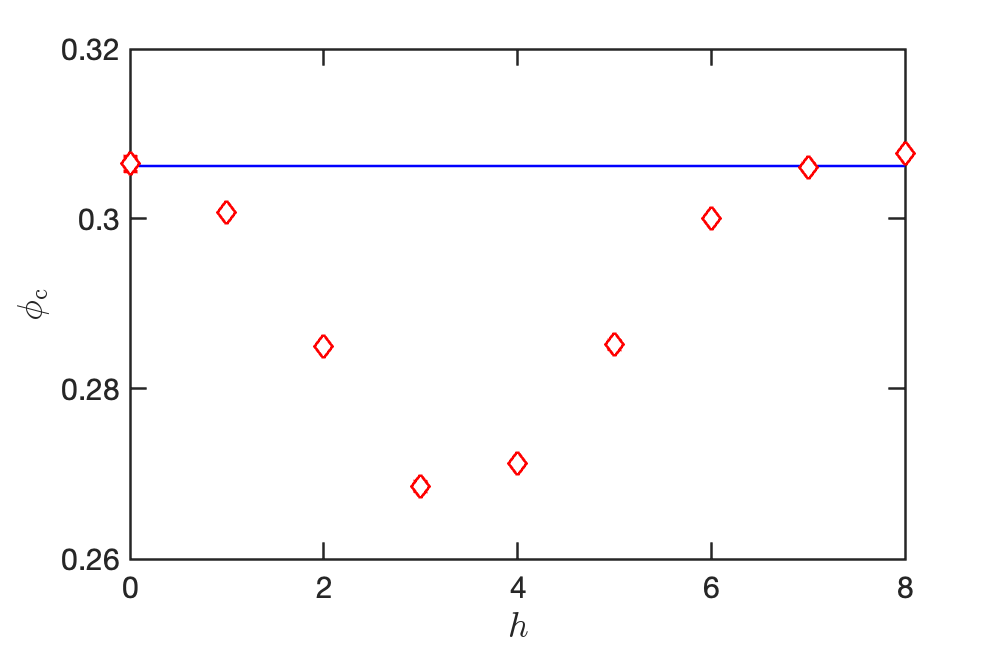}
    \caption{Critical volume fraction, $\phi_c $, as a function of the solvophilic segment length $h$, for diblock chains H$_h$T$_{r-h}$ with  overall length $r=8$. The horizontal line is the expected value based on the chain length.  Statistical uncertainties are comparable to symbol size.} 
    \label{fig:phic_db}
\end{figure}

The critical volume fraction also depends strongly on the length of the solvophilic segment, $h$, as shown in Fig. \ref{fig:phic_db}. The minimum here occurs at  $h=3$ and is approximately 12\% lower than the critical volume fraction for the homopolymer, once more a rather strong effect. As observed in Ref. \citenum{Sta20}, lower volume fractions are associated with structured liquids. For $h=7$, the critical volume fraction  recovers the value for the homopolymers ($h=$0 or 8). Note that independent simulations were done for the two homopolymers H$_8$ and T$_8$, explaining the small difference in critical volume fractions at the two ends.  The fitted line from Fig. \ref{fig:phic0} is also not a perfect match of the limiting values. Here, there is competition between two opposing effects: high values of $h$ mean that the critical temperature is lower, as seen in Fig. \ref{fig:Tc_db}. At lower temperatures the chains pack more closely, thus increasing the critical volume fraction. On the other hand, at low values of $h$, the effect of the critical temperature is  less and the behavior is dominated by the less efficient packing of T segments together because of the intervening H segments. 

\subsection{Overall Length Dependence with Fixed End Segment}

\begin{figure}
    \centering
    \includegraphics[width=3.5 in]{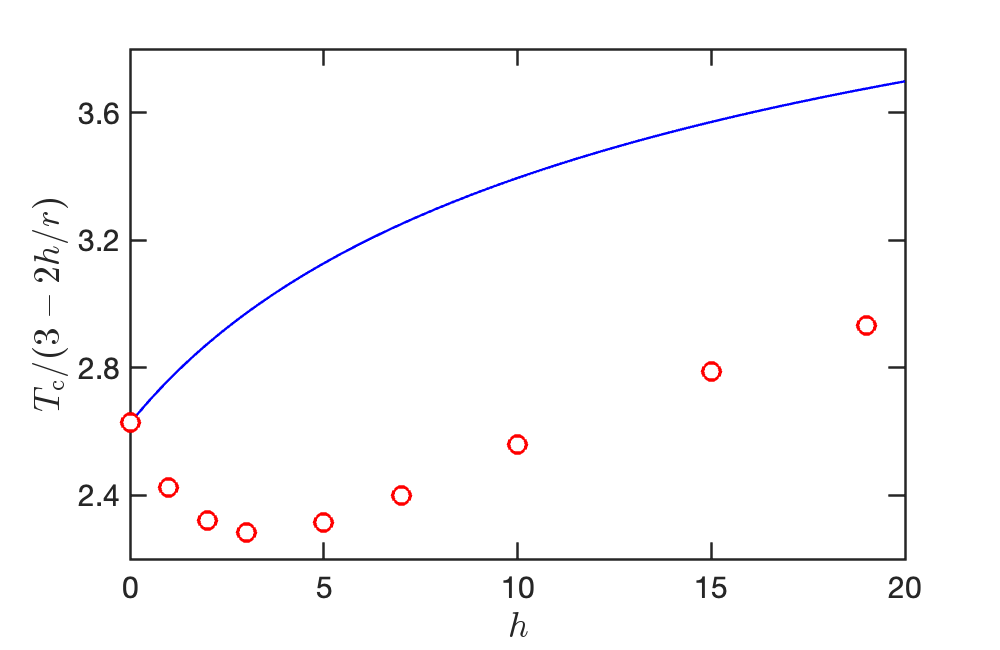}
    \caption{Normalized critical temperature, $T_\mathrm{c} / (3-2h/r)$ as a function of the solvophilic segment length $h$, for diblock chains of type H$_h$T$_5$. The line is the expected dependence based on the overall chain length. Statistical uncertainties are smaller than symbol size.} 
    \label{fig:Tc31}
\end{figure}

The next set of simulations were done for chains of variable solvophilic block length $h$, but with a fixed solvophobic end-segment length of 5 beads. The architecture is now H$_h$T$_5$ and the overall chain length is $r=h+5$. Fig. \ref{fig:Tc31} shows the measured dependence of the critical temperature as points. In order to normalize the critical temperature for the varying chain composition, we use again the scaling $T_\mathrm{c} / (3-2h/r)$.  However, in this case, the overall length of the chain is not fixed, so the appropriate comparison for the critical temperature is the length-dependent value obtained through Eq. \ref{eq:Tc}, which is shown as a blue curve in Fig. \ref{fig:Tc31}. At $h=0$, the critical temperature matches that for the homopolymer chain of length 5. Addition of solvophilic beads to the chain results in a rapid decrease of the critical temperature, since the added H beads ``protect'' the solvophobic T beads from coming to contact as readily and thus lower the effective interactions. The critical temperature gets lower, even though the expectation based on the overall chain length would be for the critical temperature to increase. At the opposite limit of large $h$, the presence of a relatively short solvophobic group at the end of the chain has a diminishing overall effect, so it is expected that the data will approach asymptotically the homopolymer result. Indeed, the data points and the line in Fig. \ref{fig:Tc31} become parallel to each other. Convergence between them would happen for much longer chains that the ones studied here; for $r=24$ there is still a sizable gap between the actual critical temperature and that expected based on the chain average composition and overall length. The total effect on the normalized critical temperature results in a minimum around $h=4$, similar to that observed in the previous subsection for diblock chains of fixed overall length.  

\begin{figure}
    \centering
    \includegraphics[width=3.5 in]{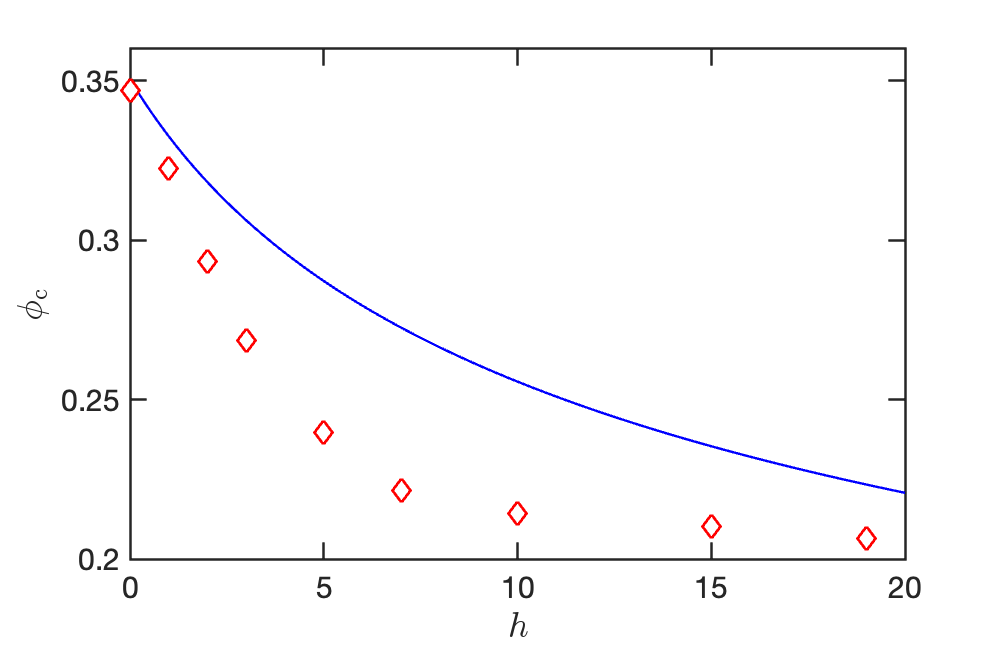}
    \caption{Critical volume fraction, $\phi_c $ as a function of the solvophilic segment length $h$, for diblock chains of type H$_h$T$_5$. The line is the expected dependence based on the overall chain length. Statistical uncertainties are  comparable to symbol size.} 
    \label{fig:phic31}
\end{figure}

Results for the critical volume fraction for systems with increasing solvophilic block length and a fixed solvophobic end segment are shown in Fig. \ref{fig:phic31} as points, along with the curve obtained by fitting the homopolymer critical volume fraction data of Fig. \ref{fig:phic0}. The critical volume fraction rapidly decreases, at a rate much faster than that expected on the basis of the overall chain length. The gap between the curve and the points is maximum for $7 \le h \le 10$. The critical volume fraction then plateaus and converges to the value expected based on the length, as seen on the right side of the figure. Once more, the physical explanation for this behavior is the less efficient packing of chains when a solvophilic block is added. 

\begin{figure}
    \centering
    \includegraphics[width=3.5 in]{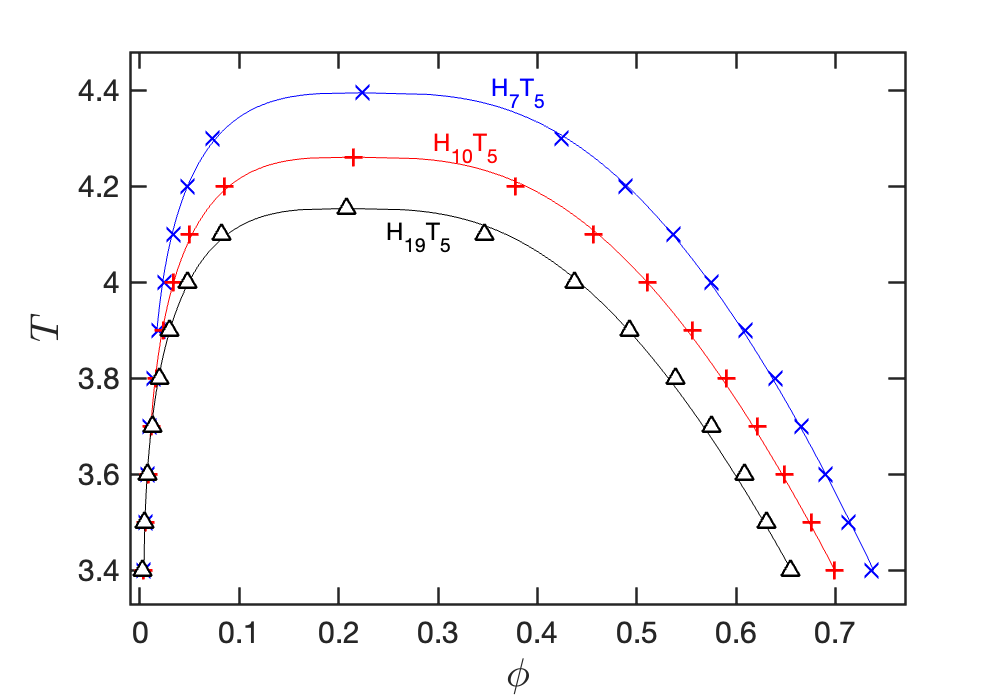}
    \caption{Temperature $T$ versus coexisting volume fraction, $\phi$, for diblock chains of type H$_7$T$_5$, H$_{10}$T$_5$, and H$_{19}$T$_5$, respectively from top to bottom. The lines are fitted to the  points as explained in the text. Statistical uncertainties are smaller than symbol size.} 
    \label{fig:phase}
\end{figure}

Calculations of the phase coexistence (binodal) curves were performed for selected systems, specifically for the architectures H$_7$T$_5$, H$_{10}$T$_5$, and H$_{19}$T$_5$, as shown on Fig. \ref{fig:phase}. The lines on this figure were obtained by fitting coexistence data below the critical point to the scaling relationship appropriate for the 3-dimensional Ising universality class,
\begin{equation} 
		\phi_\mathrm{l} - \phi_\mathrm{g} = a( T _\mathrm{c} - T)^ \beta
\end{equation}
where $\phi_\mathrm{l}$ and $\phi_\mathrm{g}$ are the volume fractions of the two phases at coexistence, $a$ is a fitted constant, and $\beta=0.326$ is the scaling exponent for the coexistence width. This expression was combined with the approximate ``law of rectilinear diameters,''
\begin{equation} 
		\phi_\mathrm{l} + \phi_\mathrm{g} =  b( T _\mathrm{c} - T) + 2\phi_\mathrm{c}
\end{equation}
where $b$ is another fitted constant. Analytical expressions were obtained for  $\phi_\mathrm{l}$ and $\phi_\mathrm{g}$ as functions of $T$, using the known values of  $T _\mathrm{c}$ and $\phi_\mathrm{c}$  for each system and separate optimizations of parameters $a$ and $b$ for $\phi_\mathrm{l}$ and $\phi_\mathrm{g}$. One of the reasons for determining these coexistence curves was to ensure that there are no finite-size aggregates (micelles) present in the low-density phase. Indeed, simulation runs up to (and even slightly beyond) the coexistence chemical potential at a range of temperatures below the critical point displayed monotonic decay of the aggregate size distributions with no micellar aggregates present. This behavior stands in contrast to that observed for comparable chain lengths and architectures in Ref. \citenum{Pan23}, because of the use of a reduced-contrast interaction set in the present study that makes finite-size aggregates less stable relative to the bulk condensed phase. 

\subsection{Single Bead Position Dependence}

\begin{figure}
    \centering
    \includegraphics[width=3.1 in]{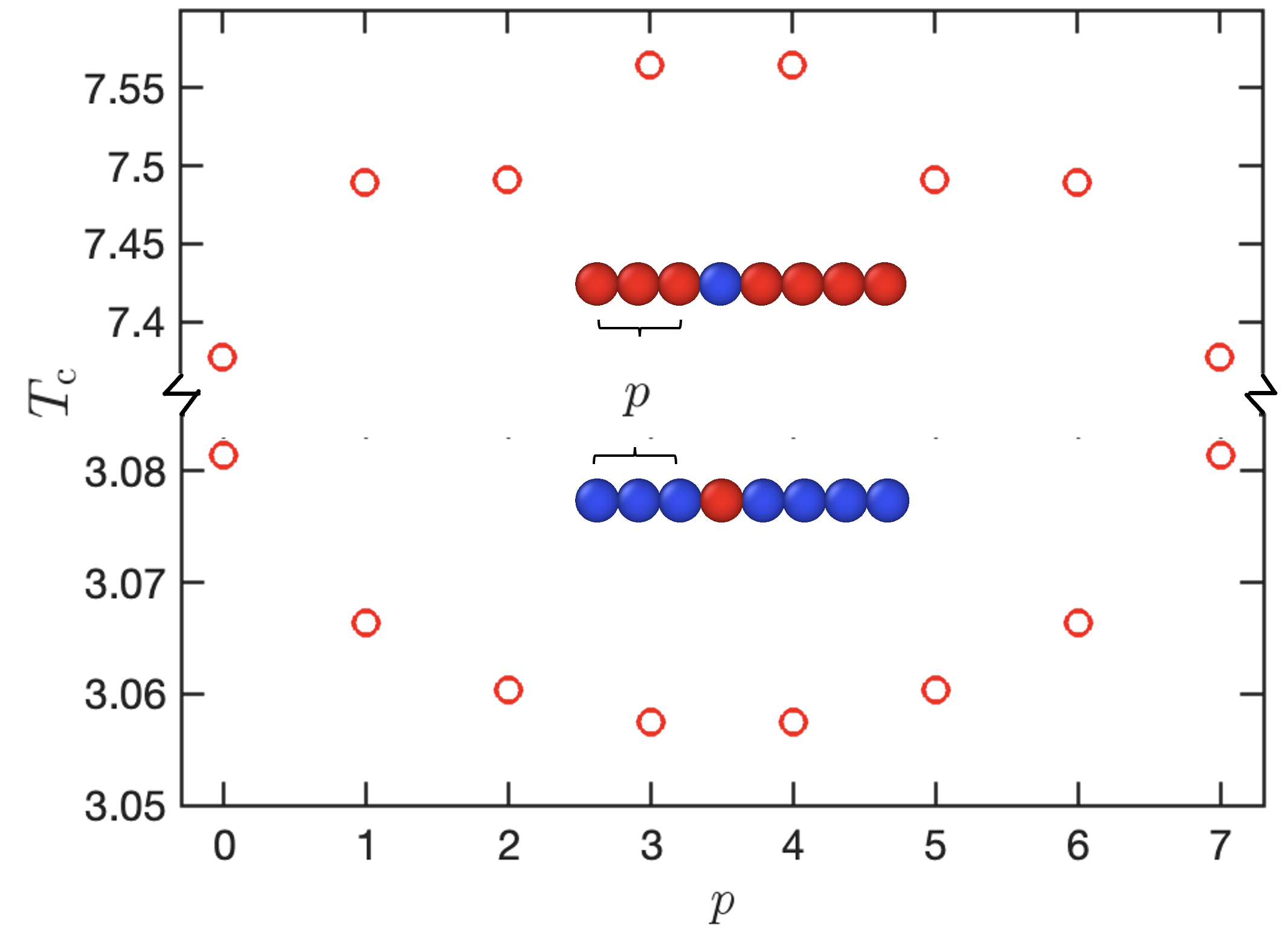}
    \caption{Critical temperature, $T_c$, as a function of the distance from the end, $p$, of a single ``different'' bead in chains with overall length $r=8$. The general structures are H$_p$TH$_{7-p}$ (bottom) and T$_p$HT$_{7-p}$  (top). Data have been reflected with respect to the midpoint of the chain due to symmetry. Statistical uncertainties are smaller than symbol size.} 
    \label{fig:rm}
\end{figure}

Additional calculations were performed for chains of length $r=8$ that consisted of beads of the same character, except for a single ``different'' bead placed at variable locations, from the end to the middle of the chain. Thus, the architectures were H$_p$TH$_{7-p}$ or T$_p$HT$_{7-p}$, where $p$ is the number of beads prior to the position of the bead of different character and can range between 0 and $r-1$. Results for the critical temperature of such systems are shown in Fig. \ref{fig:rm}. As seen in the bottom part of the figure, moving a solvophobic bead from the end of the chain  to the middle where it can less readily interact with beads in other chains, results in a slight decrease of the critical temperature, of around  $-0.8$\%. Conversely, as seen in the top part of the figure, moving a solvophilic bead from the end of a chain to the middle, where is less disruptive to solvophobic bead attractions across chains, results in an increase in the critical temperature, of significantly higher magnitude, around $2.5\%$ for this system.  For comparison, in Ref. \citenum{Sta20}, a 5\% increase in critical temperature was seen for chains of length $r=20$ when a single solvophilic bead was moved from the end  to the middle of an otherwise solvophobic chain. Thus, the magnitude of the effect seems to be comparable, even though different models were used in the two studies (continuum versus lattice model). 

The effects on the critical volume fraction of moving a single bead of opposite character along a chain were quite small, within statistical uncertainties of the corresponding value in almost all cases. These values are listed in the additional material provided online, as explained in the Data Availability Statement. 

\subsection{Overall Architecture Effects}
\begin{figure}
    \centering
    \includegraphics[width=3.5 in]{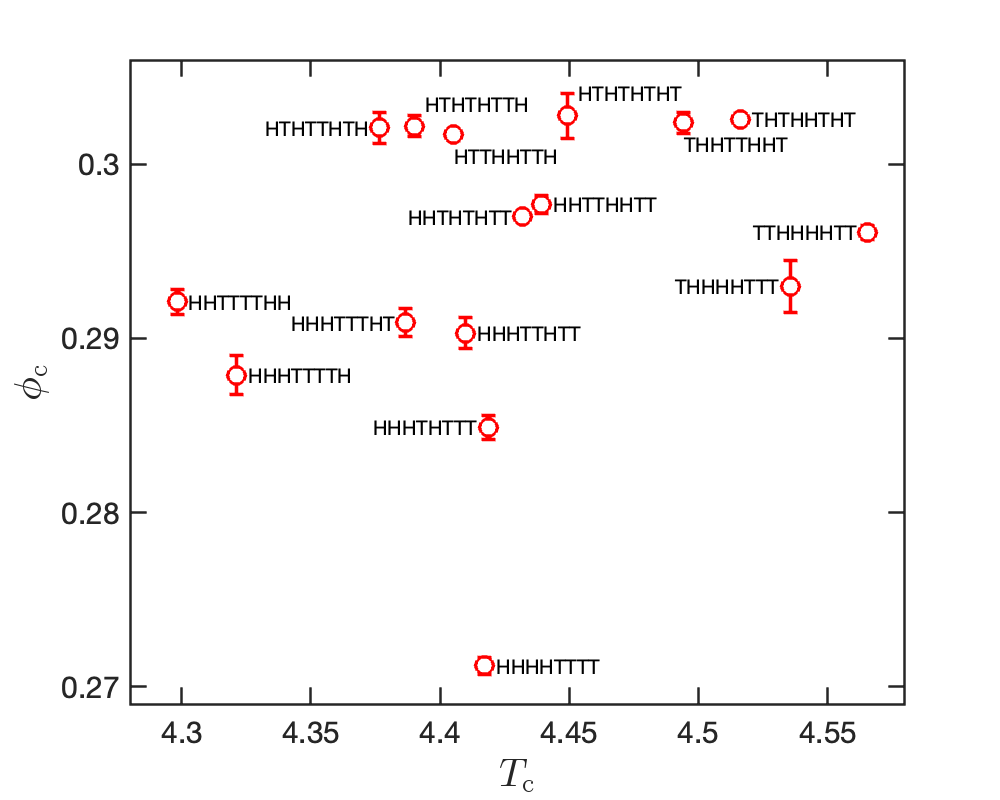}
    \caption{Critical volume fraction, $\phi_c $ as a function of the critical temperature, $T_c$, for sequences with 50\% H groups, $r=8$. Statistical uncertainties for the critical volume fractions are shown when larger than symbol size. Statistical uncertainties for the critical temperatures are smaller than symbol size.} 
    \label{fig:Tcpc}
\end{figure}

The final set of comparisons was made using a variety of chain architectures for chains with $r=8$, all of the same composition, specifically 50\% solvophobic. There are 38 distinct sequences of 8-mers with 4 T beads and 4 H beads, when one takes into account front-to-back symmetric sequences. Fig. \ref{fig:Tcpc} displays the critical volume fraction plotted against the critical temperature for a selected subset of the possible sequences for such chains. There is a good amount of spread in both critical temperatures and critical volume fractions of the different architectures. Certain features emerge clearly: for example, the lowest by far critical volume fraction corresponds to the diblock sequence, HHHHTTTT. On the other hand, this sequence has an unremarkable critical temperature in the ``middle of the pack.'' The tendency of more blocky sequences to possess lower critical volume fractions was also observed in Ref. \citenum{Ran21}, where it was determined that sufficient blocky sequences form finite-size micelles rather than phase separating, because of the full-contrast interaction set used in the prior study.

The highest critical temperatures is observed here for the symmetric triblock sequence, TTHHHHTT, in which the solvophobic beads are at the chain ends, thus available to interact with solvophobic beads in other chains. Conversely, the symmetric triblock sequence, HHTTTTHH, has the lowest critical temperature of all architectures, since it has ``burried'' solvophobic groups away from the chain ends. This pattern is seen more generally -- H groups at the chain ends are associated with low critical temperatures, whereas T groups with higher. Chains with alternating bead types (e.g., HTHTHTHT), or similar (e.g., THTHHTHT) have the highest critical volume fractions, suggesting more efficient packing in the condensed state.  

\section{Conclusions}
This work focused on the accurate determination of critical properties for short, oligomeric chains consisting of two types of beads, solvophobic and solvophilic. A previously proposed lattice model was used, with parameters modified to prevent formation of finite-size clusters that disrupt phase transitions into bulk macroscopic liquid phases. While conceptually simple, the model reproduces many features of the experimentally determined behavior of polymer/ solvent systems, specifically with respect to the scaling of critical parameter with chain length.\cite{Dob80}  The critical temperature and critical volume fraction pass through a minimum when a solvophilic segment is added to a solvophobic block of a chain. The location of this minimum with respect to the size of the solvophilic block is different for the two properties, reflecting differences in the physical mechanisms responsible for the reduction in the critical temperature versus the critical volume fraction. The position of a single bead of solvophilic or solvophobic character within a chain composed of opposite-type monomers leads to a maximum or minimum, respectively, in the critical temperature. There is no appreciable change in the critical volume fraction for these systems. When examining a range of architectures at a fixed overall chain length and composition, there are significant variations of the critical temperature and volume fraction. Architectures with solvophobic beads near the chain ends have higher critical temperatures and diblock chains have the lowest critical volume fraction. 

One open question that could be addressed in future studies is the character of the transition from simple phase behavior (with no aggregation) observed in the current study that used the reduced-contrast interaction set, with the observed formation of finite-size aggregates seen in previous studies of related models with a higher-contrast interaction set. Most likely, for specific architectures and interaction parameters, one should be able to observe both phase separation and aggregation in the same system, as seen in a prior computational study\cite{Pan23} and also experimentally.\cite{Tay24}  The scaling of the location of the transition with respect to chain length, composition, and strength of interactions would be of interest.

The present work has implications on the design of synthetic polymers with desired phase separation properties and in understanding biomolecular phase separation. It could also be useful in providing data for the development of refined theoretical models for heteropolymer and intrinsically disordered protein phase behavior. These models could be based on molecularly informed field theoretic\cite{Ngu23} or other variational methods.\cite{Saw15} Another possibility would be to utilize machine-learning methods (e.g.,\cite{Tan24}) to correlate chain architecture and the phase behavior. This latter approach would require much larger datasets for training the machine learning models than what was generated in the present work, but this should be computationally feasible with the methods used here. 

\section*{Acknowledgements}

Financial support for this work was provided by the Princeton Center for Complex Materials (PCCM), a U.S. National Science Foundation Materials Research Science and Engineering Center (Award DMR-2011750). 

\section*{Data Availability Statement}
Computer codes used in this work, example input and output files, information on the runs performed, and numerical data for the critical points and phase coexistence volume fractions are freely available for download from the Princeton Data Commons repository, at DOI 10.34770/2xrn-jt41. 

\section*{References}
\bibliography{library}

\end{document}